\documentclass[aps,prl,twocolumn,groupedaddress]{revtex4}

\usepackage{amssymb}
\usepackage{amsmath}
\usepackage{bm}
\usepackage{graphicx}
\usepackage{dcolumn}
\usepackage{longtable}
\usepackage{color}

\begin{document}


\title{Microscopic coexistence of superconductivity and charge order in organic superconductor $\beta''$-(BEDT-TTF)$_{4}$[(H$_3$O)Ga(C$_2$O$_4$)$_3$]$\cdot$C$_6$H$_5$NO$_2$}


\author{Y.~Ihara}
\affiliation{Department of Physics, Graduate School of Science, Hokkaido University, Sapporo 060-0810, Japan}
\email{yihara@phys.sci.hokudai.ac.jp}
\author{K.~Moribe}
\affiliation{Department of Physics, Graduate School of Science, Hokkaido University, Sapporo 060-0810, Japan}
\author{S.~Fukuoka}
\affiliation{Department of Physics, Graduate School of Science, Hokkaido University, Sapporo 060-0810, Japan}
\author{A.~Kawamoto}
\affiliation{Department of Physics, Graduate School of Science, Hokkaido University, Sapporo 060-0810, Japan}


\date{\today}

\begin{abstract}
The electron paramagnetic resonance study for an organic superconductor $\beta''$-(BEDT-TTF)$_{4}$[(H$_3$O)Ga(C$_2$O$_4$)$_3$]$\cdot$C$_6$H$_5$NO$_2$ 
reveals that superconductivity coexists uniformly with the charge ordered state in one material. 
In the charge ordered state, the interplane spin exchange is gapped, while the in-plane conductivity is not significantly modified. 
This anisotropic behavior is explained by the exotic charge ordered state, in which molecular-site selective carrier localization coexists with 
conducting carriers on other molecules. 
Relationship between superconductivity and this conductive charge ordered state is investigated. 
\end{abstract} 

\maketitle

The metal-insulator transition in a partially filled conduction band is caused by the strong electron-electron interactions. 
In the case of a half-filled band, the strong on-site Coulomb repulsion prevents the carriers from moving to stabilize the Mott-Hubbard insulating state. 
While in a quarter-filled band, the long-range Coulomb interaction is responsible for the charge ordered (CO) insulating state. 
Superconductivity sometimes appears in the vicinity of the CO state, \cite{kurmoo-SM27, kobayashi-CL15, mori-PRB57}
leading us to address that the long-range electron-electron interactions can be the source of the superconducting (SC) pairing interaction 
in the same manner as the on-site interaction contributes to form unconventional superconductivity near the magnetic instability. \cite{mathure-Nat394} 
Since many unconventional superconductors have been found near the magnetic critical points, 
we naturally expect that intriguing SC features would be found for superconductivity near the charge instability. 
However, experimental realization is difficult because the charge ordering coincides with the strong carrier localization, 
which severely conflicts with the itinerant nature of the SC state. 
Even though, a theoretical study based on the extended Hubbard model shows that 
the charge fluctuations near the charge ordering transition can contribute to the SC pair formation
when the charge ordering transition is suppressed to very low temperatures. \cite{merino-PRL87}
In fact, the increase in the charge fluctuations is observed in the organic superconductor from an optical study, \cite{kaiser-PRL105}
suggesting the interplay between the charge instability and superconductivity. 
Further experimental identification for the relationship between the neighboring, 
and even coexisting CO and SC states is crucial to expand the possibility of SC pairing mechanism. 

To investigate experimentally the effect of long-range electron-electron interactions on superconductivity, 
layered organic superconductors are the best suited because of their low carrier density, and thus the weak screening effect. 
Several organic conductors, such as $\alpha$-type and $\theta$-type BEDT-TTF salts (BEDT-TTF: bisethyldithiotetrathiafulvalen) show 
apparent charge ordering transitions. \cite{mori-PRB57, kakiuchi-JPSJ76, nogami-SM103, miyagawa-PRB62}
Theoretical studies for $\theta$-type salts proposed a three-fold CO state, \cite{mori-JPSJ72}
which can maintain the metallic conductivity even when some part of the carriers are localized at one of three molecular sites. 
Further theoretical studies show that when the itinerant carriers interact with the localized charges, 
unconventional metallic state referred to as the pinball liquid state will be realized. \cite{kaneko-JPSJ75, hotta-JPSJ75, merino-PRL96}
Experimentally, however, such three-fold CO state has not been found in $\theta$-type salts, \cite{watanabe-JPSJ73}
because the three-fold CO state appears only at a limited parameter space, 
where the stability of the horizontal stripe phase competes with the vertical phase. \cite{mori-JPSJ72, kaneko-JPSJ75}
To reveal the novel CO state and its impact to the SC pair formation,
we should study materials which show the non-trivial CO state very close to the SC transition temperature. 

A layered organic superconductor $\beta''$-(BEDT-TTF)$_{4}$[(H$_3$O)Ga(C$_2$O$_4$)$_3$]$\cdot$C$_6$H$_5$NO$_2$ ($\beta''$-Ga) salt \cite{akutsu-JACS124}
is an ideal material to explore the charge order and superconductivity, 
because this compound shows charge instability at a temperature very close to the SC transition temperature $T_{c} = 7$ K. 
Previous NMR and EPR experiment detected no anomaly due to magnetic phase transition at the charge ordering temperature $T_{\rm CO} = 8.5 K$, \cite{ihara-JPSJ82, ihara-JPSJ85}
which was determined as the onset of the NMR spectrum splitting. \cite{ihara-PRB90}
Also, the NMR experiments showed that the low-energy spin dynamics increases at $T_{\rm CO}$, 
which is suggestive of the increase in the charge fluctuations near the SC transition. \cite{ihara-JPSJ82}
Besides, from the NMR intensity ratio between the charge rich and poor sites, the three-fold charge pattern has been suggested. \cite{ihara-PRB90}
As the charge localization in the CO state coexists with the metallic conductivity, and even with superconductivity below $T_{c}$, 
$\beta''$-Ga salt is one of the best candidates for the experimental realization of the pinball liquid state. 
However, as the resistivity experiments suggest a possibility of phase segregation, that is, the CO part of the sample is separated from the SC part, \cite{coldea-PRB69}
we should clarify if the CO state coexists microscopically with superconductivity or not. 
The NMR spectroscopy is one of the most powerful technique to investigate the electronic state from a microscopic viewpoint. 
However, because of the insufficient spectrum resolution, 
we were not able to exclude the possibility of phase segregation 
An alternative probe with higher resolution was desired. 

In this study, we show that the X-band electron paramagnetic resonance (EPR) experiment is one such probe. 
The EPR signal in $\beta''$-Ga salt originates from the $\pi$ electrons in the highest occupied molecular orbital of the BEDT-TTF molecules.
We succeed in detecting the charge anomaly on the clearly resolved EPR spectrum 
by taking advantage of the anisotropy of $g$ factors and the bilayer crystal structure of $\beta''$-Ga salt. 
Thus, the EPR experiment allows us to observe the phase segregation, if any, as the additional component of the EPR spectrum. 
The present results, which are explained by a single EPR contribution at any temperatures, clearly evidence the uniform coexistence between SC and CO states. 
We also conducted the resistivity measurement to confirm that the CO state actually involves the conducting carriers, 
because the EPR experiment is in principle a spin sensitive probe. 
The results of resistivity and EPR measurements are explained consistently, which unambiguously suggest a SC state coexisting with the conductive CO state.

\begin{figure}[tbp]
\begin{center}
\includegraphics[width=8.5cm]{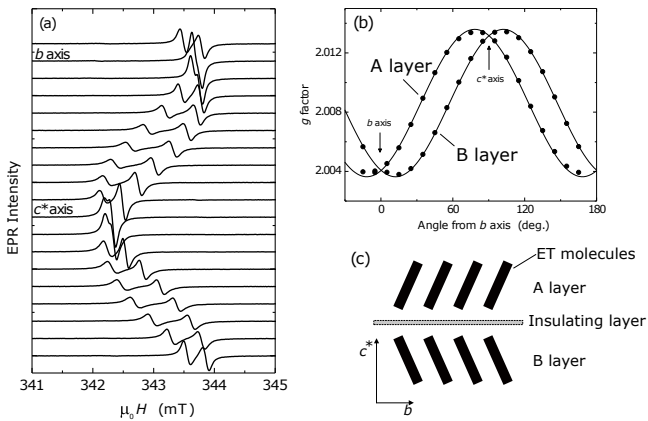}
\end{center}
\caption{
(a) X-band EPR spectra of $\beta''$-Ga salt at $3.6$ K in various external field directions. 
The two-peak spectrum observed in arbitral field direction merges into a single peak in fields along the symmetric axes. 
(b) Field direction dependence of the $g$ factors determined from the EPR peak positions. 
Two branches with sinusoidal angle dependence originate from the A and B layers. 
(c) The double-layer crystal structure of $\beta''$-Ga salt. 
The conducting BEDT-TTF molecules layers are separated by the insulating anion layers. 
The long axis of the BEDT-TTF molecules in layer A (B) is inclined by $76^{\circ}$ ($-76^{\circ}$) with respect to the $b$ axis, 
which coincides with the maximum in the angle dependence of $g$ factor (b). 
}
\label{fig1}
\end{figure}

The single crystalline samples were grown by the standard electrochemical reaction. \cite{akutsu-JACS124}
The X-band EPR experiments were performed with a commercial spectrometer (Bruker EMX Plus). 
One single crystal with a dimension of $2\times 0.5\times 0.2$ mm$^3$ was used for the experiment.  
The orientation of the external magnetic field was tuned by the single axis rotator, 
with which the sample can be rotated {\it in situ} around the crystalline $a$ axis. 
We also performed the in-plane ($b$ axis) and interplane ($c^\ast$ axis) resistivity measurements by the conventional 4 probe method. \cite{sup}

Figure~\ref{fig1} (a) shows the EPR spectra at the lowest temperature of $3.6$ K in the $bc^\ast$ plane fields. 
The field direction is determined by the angle from the $b$ axis. 
The angular dependence of the EPR spectrum originates from the anisotropic $g$ tensor, 
for which the principal axes coincide with the symmetric axes of the BEDT-TTF molecule, 
and the principal values were determined for $\beta$-(BEDT-TTF)$_2$I$_3$ salt as $g=(2.011, 2.008, 2.002)$. \cite{kinoshita-JPSJ54}
The $g$ factors for each field direction were determined by fitting the spectra with two-peaks Lorenzian function. 
As the result, two sinusoidal branches were obtained as shown in Fig.\ref{fig1}(b). 
These branches are assigned to the EPR signals from the A and B layers displayed in Fig.~\ref{fig1} (c). 
Assuming that the principal axes of $g$ tensor are fixed to the BEDT-TTF molecules, 
we can determine the principal values of $g$ tensor for $\beta''$-Ga salt as $g=(2.016, 2.010, 2.001)$. 
Good agreement with previous study \cite{kinoshita-JPSJ54}  suggests that the entire EPR spectra can be explained by 
the anisotropic $g$ tensor and the double-layer crystal structure of $\beta''$-Ga salt, 
meaning that our crystal is single phase and the electronic state is uniform.

\begin{figure}[tbp]
\begin{center}
\includegraphics[width=7cm]{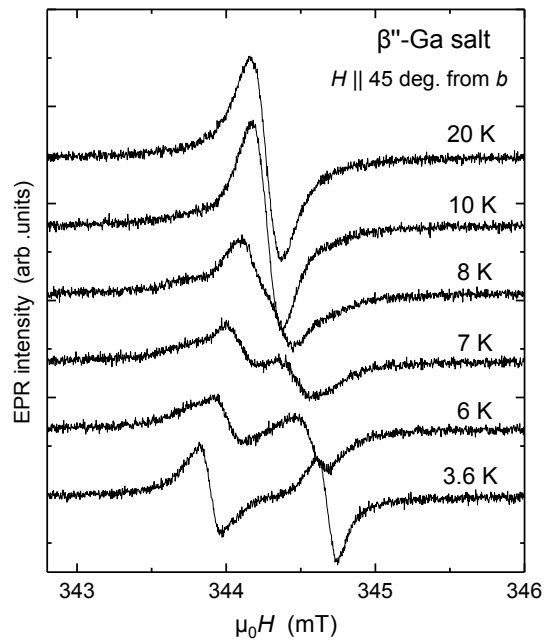}
\end{center}
\caption{
The EPR peak splitting observed below $T_{\rm CO}=8.5$ K in the $45^{\circ}$ field, 
where the peak separation at the lowest temperature becomes the maximum. 
The spectrum splitting without the increase in the spectrum width indicates the anomaly in the interplane cross relaxation (see text) 
with negligible modification in the in-plane spin relaxation. 
}
\label{fig2}
\end{figure}

\begin{figure}[tbp]
\begin{center}
\includegraphics[width=8cm]{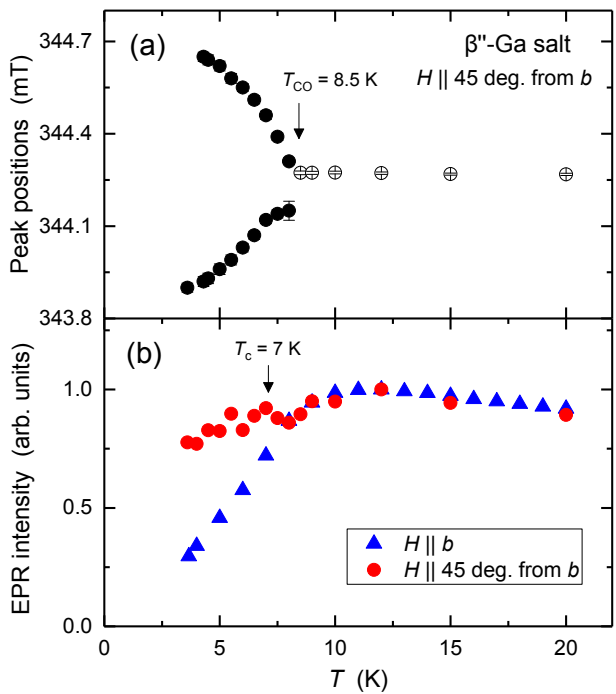}
\end{center}
\caption{
(a) The temperature dependence of the EPR peak fields. 
Abrupt peak splitting was observed below $T_{\rm CO}=8.5$ K. 
(b) The temperature dependence of the EPR intensity in the $b$-axis field, and $45^{\circ}$ field. 
The intensity is normalized at $12$ K to correct the intensity difference due to the sample shape. 
The SC transition is detected as the reduction of EPR intensity below $T_{c}$ in the $b$ axis field. 
}
\label{fig3}
\end{figure}

Next, we measured the temperature dependence of the EPR spectrum in the field applied to the direction $45^{\circ}$ rotated from the $b$ axis ($45^{\circ}$ field). 
In this field direction, $T_c$ is suppressed below $3$ K by a field of approximately $300$ mT. 
With increasing temperatures, the peak separation becomes small, as shown in Fig.~\ref{fig2}, 
and a single peak was observed at the temperatures higher than $8.5$~K. 
We found a trace of the two-peak structure at $8$ K as the wiggle around the center of the spectrum at $344.4$~mT. 
Therefore, the peak positions were determined by the two-component Lorentzian fit for the spectra below $8$ K (filled symbols in Fig.\ref{fig3} (a)), 
and by the single component Lorentzian fit above $8.5$ K (open symbols in Fig.\ref{fig3} (a)). 
The abrupt increase in the peak separation below $8$ K clearly evidences a phase transition. 
This anomaly agrees with the charge ordering transition at $T_{\rm CO}=8.5$ K previously observed from the $^{13}$C NMR study. \cite{ihara-PRB90}
The origin of the EPR peak splitting will be discussed later. 

As the charge ordering anomaly is successfully observed in the EPR spectrum, then 
we compare the EPR spectra in $b$-axis and $45^{\circ}$ fields to unveil the relationship between the CO and SC states. 
Typical EPR spectra for $b$-axis fields are presented in the supplimental material. \cite{sup}
In the $b$-axis field of $345$ mT, $T_c$ does not change because of the extremely high upper critical field ($B_{c2}>30$~T). 
The effect of SC transition was observed in the EPR spectrum as the reduction of the integrated intensity below $7$ K (Fig.~\ref{fig3} (b)).  
Such decrease in intensity was not observed in the $45^{\circ}$ field, because $T_{c}$ is suppressed below $3$ K. 
This result confirms that the electronic spins that would show superconductivity in zero field contribute to the EPR intensity when superconductivity is suppressed by $45^{\circ}$ field. 
Thus, if SC part of the sample did not show the charge ordering transition, which is the case for the macroscopic phase segregation, 
an additional EPR peak originating from the electrons in a normal metallic state should be observed at the center of the two-peak spectrum. 
However, such extra contribution was not observed at the lowest temperature of $3.6$ K, as shown at the bottom of Fig.~\ref{fig2}. 
The clear two-peak spectrum in the $45^{\circ}$ field allows us to conclude that the SC state coexists uniformly with the CO state. 
We note that the EPR intensity decreases gradually below $T_{c}$ in the $b$-axis field, and finite intensity remains even at $3.6$ K. 
This behavior is contrasting to the conventional behavior expected for a homogeneous SC state, in which EPR signal should disappear. 
The EPR in the SC state may originate from the nearly localized electrons in the CO state, 
for instance, the pin site in the pinball liquid state. 
We exclude the possibility that the entire EPR intensity originates from the free spins induced by impurity or defects, 
because the EPR intensity increases from $20$ K up to room temperature, 
which is contrast to the Curie behavior expected for the free spins. \cite{sup} 
In fact, a typical Curie-Weiss type temperature dependence was observed above $T_{c}$ in the C and Al-doped MgB$_2$, \cite{bateni-APL105, bateni-JAP117, bateni-APL108}
in which the EPR signal originates from the defects and/or impurities. 

\begin{figure}[tbp]
\begin{center}
\includegraphics[width=8cm]{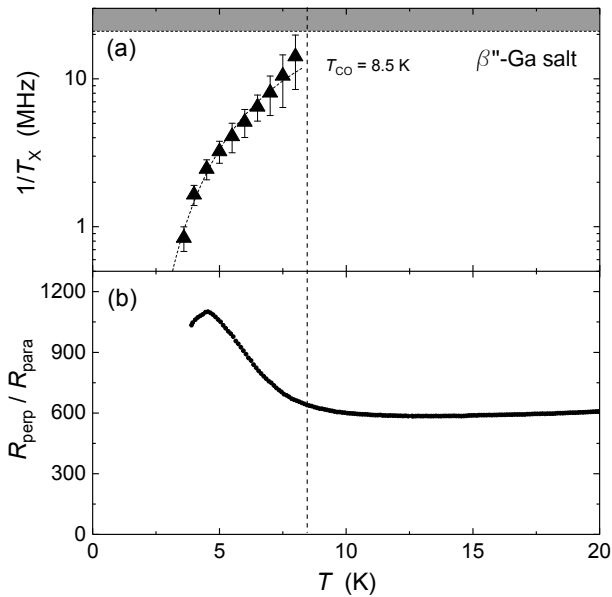}
\end{center}
\caption{
The temperature dependence of (a) the cross relaxation rate $1/T_{\rm X}$, and (b) the ratio between the in-plane and interplane resistance. 
The gray area in (a) represents the frequency that corresponds to the EPR peak separation at the lowest temperature, 
above which $1/T_{\rm X}$ cannot be determined because of the motional narrowing. 
At low temperatures, $1/T_{\rm X}$ follows an exponential function with a gap energy of $16$ K as shown by the dotted line. 
}
\label{fig4}
\end{figure}

Now, to study in detail the origin of the EPR peak splitting in the CO state, 
we fit the EPR spectra using the Bloch model with cross spin relaxation between the neighboring layers A and B (Fig.~\ref{fig1}(c)). \cite{antal-PRB84}
The time evolution of the electron magnetization on the weakly coupled layers A and B, ${\bm M}^{\rm A}$, ${\bm M}^{\rm B}$ are written as 
\begin{align}
\frac{d{\bm M}^{\rm A}}{dt} &= \frac{g^{\rm A} \mu_B}{h} \left( {\bm M^{\rm A}} \times {\bm B} \right)- {\bm R}^{\rm A} + \frac{\Delta {\bm M}^{\rm AB}}{T_X}, \\
\frac{d{\bm M}^{\rm B}}{dt} &= \frac{g^{\rm B} \mu_B}{h} \left( {\bm M^{\rm B}} \times {\bm B} \right)- {\bm R}^{\rm B} + \frac{\Delta {\bm M}^{\rm BA}}{T_X}, 
\end{align}
where ${\bm R}^{\alpha} = (M_x^{\alpha}/T_2, M_y^{\alpha}/T_2, (M_z^{\alpha}-M_0)/T_1)$ with $\alpha = $ A, B layers, 
$ \Delta {\bm M}^{\rm AB} = {\bm M}^{\rm B} - {\bm M}^{\rm A}$, and 
$g^{\alpha}$, $\mu_B$, $h$ are the $g$ factor for the $\alpha$ layer, the Bohr magneton and the Plank's constant. 
When the intrinsic spin relaxation time $T_{2}$ becomes short, the whole spectrum is broadened in proportion to $1/T_{2}$, 
and the broad spectrum smears out the two-peak structure. 
Contrastingly, when the cross relaxation time $T_{X}$ becomes faster than the timescale equivalent to the spectrum separation, 
the two-peak structure is lost due to the dynamical narrowing effect, and a single sharp peak will be observed. 
The sharp EPR spectra at high temperatures above $8.5$ K in the present study is explained by this dynamical narrowing effect. 
Figure \ref{fig4} (a) shows the temperature dependence of $T_{X}$ determined by fitting the experimental spectra with the coupled Bloch model. \cite{sup}
The gray region in Fig.~\ref{fig4} (a) represents the timescale shorter than the spectrum separation of approximately $20$ MHz.

In the CO state, 
the temperature dependence of $1/T_{X}$ shows an exponential behavior with a gap energy of $\Delta /k_B=16$ K. 
This gap size is consistent with that expected for the weak coupling charge density wave state, $2\Delta/k_BT_{\rm CO}=3.8$. 
The charge gap was detected in the interplane spin exchange channel because 
the already small interplane transfer integral in the metallic state is completely lost in the CO state due to the partial charge localization. 
Whereas for the in-plane spin relaxation $T_{2}$, we found that the spectrum width is invariant above and below $T_{\rm CO}$, 
which leads us to conclude that $T_2$ is not significantly modified by the charge ordering transition. 
The invariant $T_2$ allows us to exclude the possibility of magnetic phase transition.
This anisotropic behavior is consistently explained by the three-fold CO state, 
in which partial carrier localization immediately extinguishes the weak interplane transfer channel, 
while maintaining the coherent in-plane transfer channel. 

The anisotropy developing in the CO state is also observed from the electrical transport experiment, as shown in Fig.~\ref{fig4}~(b). \cite{sup}
The in-plane resistance ($R_{\rm perp}$) and interplane resistance  ($R_{\rm para}$) were measured in magnetic fields of $0.3$ T and $2.5$ T applied perpendicular to the conducting plane 
to suppress $T_c$. 
The increase in the ratio $R_{\perp}/R_{||}$ below $T_{\rm CO}$  means that the interplane resistivity increases due to the gap opening in the interplane transfer channel. 
A kink was observed at a lower temperature of $4.5$ K because of the precursor to the SC transition. 
The anomaly at $T_{\rm CO}$ was not clearly observed in the previously reported in-plane resistivity measurements, \cite{akutsu-JACS124, ihara-JPSJ82} 
because the highly conducting in-plane channel is not significantly modified in the CO state, which is consistent with the invariant $T_2$. 
The highly conducting in-plane channel is also suggested from the quantum oscillation study, 
in which clear Shubnikov-de Haas oscillation was observed even in the CO state, 
where resistivity shows a semiconducting temperature dependence. \cite{coldea-PRB69, bangura-PRB72}
Observation of the Fulde-Ferrell-Larkin-Ovchinnikov superconducting state in high magnetic fields is anothor evidence
for a clean electronic state in the conducting plane. \cite{uji-PRB97}

Finally, a question is whether the CO state supports or suppress superconductivity. 
In a series of $\beta''$ type BEDT-TTF salts with various guest molecules and metallic ions, 
$T_c$ higher than $5$ K is found only in the salts with resistivity upturn at low temperatures. \cite{bangura-PRB72} 
Whereas in the $\beta''$-Rh salt with $T_c=2.5$ K, almost no resistivity upturn was observed. \cite{martin-IC56}
As the charge ordering transition increases the SC transition temperatures, 
we suggest the importance of the exotic electronic state with the spatial charge modulation \cite{hotta-JPSJ75, merino-PRL96} 
to induce this unconventional type of superconductivity. 

To conclude, 
we performed the EPR experiment for the organic superconductor $\beta''$-Ga salt, 
and detected clearly the charge ordering transition at $T_{\rm CO}=8.5$ K. 
In the CO state, the interplane spin exchange channel is gapped as the result of the partial charge localization. 
In the same sample, we observed the SC transition as the reduction of EPR intensity below $T_{c}$. 
We found only a single spectrum component both in the SC and CO states, 
which suggests a uniform coexistence between superconductivity and charge order. 
To understand this exotic coexisting state, we suggest a three-fold CO state, 
in which partial charge localization coexists with the high conductivity.

\begin{acknowledgements}
We would like to acknowledge Y. Oshima, L. Martin for fruitful discussions, 
and M. Fujiwara for the support in carrying out the EPR experiment in the Institute for Molecular Science. 
This study was partly supported by the Suhara memorial foundation.
\end{acknowledgements}

\end{document}